\documentclass[prl,amsmath,amssymb,aps,superscriptaddress,floatfix,reprint,notitlepage]{revtex4-1}
\usepackage{graphicx}
\usepackage{color}
\usepackage{dcolumn}
\usepackage{bm}
\usepackage[utf8]{inputenc}
\usepackage{amssymb}
\usepackage{color,amsmath}
\usepackage{mathtools}
\usepackage{soul,xcolor} 
\setstcolor{red}
\usepackage{epstopdf}

\newcommand{\be}{\begin{equation}}
\newcommand{\ee}{\end{equation}}

\begin{document}

\title{Cascade of multi-electron bubble phases in monolayer graphene at high Landau level filling
}

\author{Fangyuan Yang}
\affiliation{Department of Physics, University of California at Santa Barbara, Santa Barbara CA 93106, USA}
\author{Ruiheng Bai}
\affiliation{Department of Physics, University of California at Santa Barbara, Santa Barbara CA 93106, USA}
\author{Alexander A. Zibrov}
\affiliation{Department of Physics, University of California at Santa Barbara, Santa Barbara CA 93106, USA}
\author{Sandeep Joy}
\affiliation{Department of Physics, Ohio State University, Columbus, Ohio 43210, USA}
\author{Takashi Taniguchi}
\affiliation{National Institute for Materials Science, 1-1 Namiki, Tsukuba 305-0044, Japan}
\author{Kenji Watanabe}
\affiliation{National Institute for Materials Science, 1-1 Namiki, Tsukuba 305-0044, Japan}
\author{Brian Skinner}
\affiliation{Department of Physics, Ohio State University, Columbus, Ohio 43210, USA}
\author{Mark O. Goerbig}
\affiliation{Laboratoire de Physique des Solides, CNRS UMR 8502, Universit\'e Paris-Saclay, 91405 Orsay Cedex, France}
\author{Andrea F. Young}
\email{andrea@physics.ucsb.edu}
\affiliation{Department of Physics, University of California at Santa Barbara, Santa Barbara CA 93106, USA}
\date{\today}

\begin{abstract}
The phase diagram of an interacting two-dimensional electron system in a high magnetic field is enriched by the varying form of the effective Coulomb interaction, which depends strongly on the Landau level index. While the fractional quantum Hall states that dominate in the lower energy Landau levels have been explored experimentally in a variety of two-dimensional systems, much less work has been done to explore electron solids owing to their subtle transport signatures and extreme sensitivity to disorder.  Here we use chemical potential measurements to map the phase diagram of electron solid states in $N=2$, $N=3$, and $N=4$ Landau levels in monolayer graphene. 
Direct comparison between our data and theoretical calculations reveals a cascade of density-tuned phase transitions between electron bubble phases up to two, three or four electrons per bubble in the N=2, 3 and 4 Landau levels respectively. Finite temperature measurements are consistent with melting of the solids for T$\approx$1K. 
\end{abstract}

\maketitle

In an electron solid, spatial translation symmetry is spontaneously broken so that the ground state charge density forms a periodic structure incommensurate with the underlying crystal lattice. One known example is obtained in high Landau levels (LLs) in two-dimensional (2D) electron systems. Theoretically, the phase diagram is expected to host a rich interplay of competing phases\cite{koulakov_charge_1996,fogler_ground_1996,moessner_exact_1996,haldane_spontaneous_2000,shibata_ground-state_2001,beig_stripe_2002,cote_dynamics_2003,goerbig_competition_2004}. A unique feature of electron solids in higher LLs is that a variable number of electrons may cluster on each site of the emergent crystal.  The formation of the phases---known as ``electron bubbles''---is driven by the structure of the electronic form factors in the LLs.  
Electron bubble phases were first identified in the GaAs 2D electron gas by the observation of re-entrant integer quantum Hall effect (RIQHE) in transport measurement\cite{lilly_evidence_1999, du_strongly_1999}, in which the crystallized electrons freeze and no longer contribute to the Hall conductivity. Similar phases are also expected in graphene\cite{papic_tunable_2011,knoester_electron-solid_2016,zhang_wigner_2007}, and recent measurements have confirmed their existence\cite{chen_competing_2019,zeng_high-quality_2019}. 
While the existence of electron solids is straightforward to confirm using transport measurements, distinguishing them from each other to construct a comprehensive phase diagram is not.  To this end, other experimental methods, such as microwave spectroscopy\cite{lewis_microwave_2002}, surface acoustic wave transmission\cite{msall_acoustic_2015,friess_negative_2017}, and tunnelling spectroscopy \cite{jang_sharp_2016} have been developed to study vibrating modes related to the lattice structure of electron solids. More recently, temperature dependent transport has shown that the same RIQH state may host more than one bubble phase, distinguished by different melting temperatures\cite{ro_electron_2019,fu_two-_2019,ro_stability_2020}. However, a detailed phase diagram of the electron bubble phases across different LLs, long been predicted by theory, has not been conclusively established.

Measuring thermodynamic properties provides a probe of quantities directly related to the ground state energy, offering a chance to map out a complete phase diagram independent of the detailed transport phenomenology of the ground state. 
In this Letter, we use chemical potential measurements \cite{yang_experimental_2021} to construct just such a phase diagram for partially filled LLs in monolayer graphene. Our data demonstrate the existence of multiple distinct electron bubble phases characterized by different bubble sizes.  
By directly comparing our data with mean-field-theory calculations, we establish a one-to-one correlation between the filling factor and the electron bubble morphology.

Our measurement is performed in a graphene/hBN heterostructure assembled using standard dry pickup techniques\cite{wang_one-dimensional_2013}. Two graphene monolayers are separated by an hBN dielectric layer of 40nm thickness, with additional hBN dielectric and graphite gates forming a four-plate capacitor geometry.
The top graphene serves as a charge detector, which combined with a feedback loop allows us to determine changes in chemical potential of the bottom `sample' graphene accurately\cite{yang_experimental_2021}.  


\begin{figure*}[ht!]
\includegraphics[width=\textwidth]{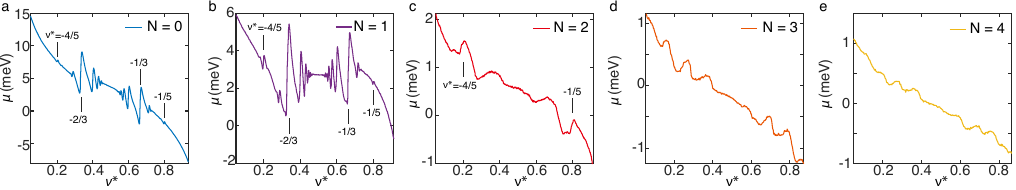}
\caption{\label{fig:figure1} 
\textbf{FQH and electron solid states in graphene monolayer probed by chemical potential measurements.} 
(a) Chemical potential change as a function of effective filling factor, $\nu^*\equiv \nu-\lfloor \nu \rfloor$, in the $N=0$, (b) $N=1$, (c) $N=2$, (d) $N=3$, and (e) $N=4$ LLs. 
In the $N=0$ and $N=1$ LLs, FQH states are observed as jumps in $\nu$ at $\nu^*=p/(2p\pm1)$ and $\nu^*=p/(4p\pm1)$ ($p=1,2,3,...$), a selection of which are labeled. For $N\geq2$, broad oscillatory features dominate, which we associated with electron solids. The $N=2$ LL is a marginal case where fractional quantum Hall states and electron bubbles compete within a narrow range of filling factors. All data measured at $B=13T$ and $T=15mK$.
}
\end{figure*}

Fig.~\ref{fig:figure1} presents the chemical potential $\mu$ measured across individual LLs with orbital quantum numbers $N=0$, 1, 2, 3, and 4.
The qualitative behavior of $\mu$ depends strongly on $N$.  For $N=0$ and $N=1$ (Figs.~\ref{fig:figure1}a-b) fractional quantum Hall states are favored, with incompressible states (manifesting here as nearly discontinuous jumps in $\mu$) observed at filling factors associated with two-flux and four-flux composite fermion sequences\cite{jain_composite-fermion_1989}. For $\nu^*>-1/5$ (or $\nu^*<-4/5$) within the $N=0$ and $N=1$ LL, $\mu$ changes smoothly, showing a large negative inverse compressibility $d\mu/d\nu$\cite{bello_density_1981}. 
This behavior has been identified with the formation of Wigner crystal states in previous experiments in both GaAs\cite{eisenstein_negative_1992,eisenstein_compressibility_1994} and graphene\cite{zhou_solids_2019,yang_experimental_2021}.

\begin{figure}[b]\vspace{-15pt} 
\includegraphics{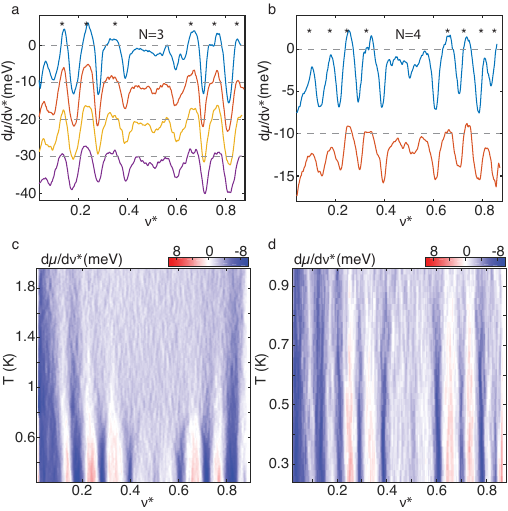}
\caption{\label{fig:figure2} \textbf{Electronic compressibility and temperature dependence of electron bubble phases. }
(a) $d\mu/d\nu$ in the $N=3$ and  (b) $N=4$ LLs. The data is obtained via numerical differentiation of $\mu$ measured at $13T$ and $15mK$. Within each LL, the four symmetry breaking levels are plotted by blue, red, orange and purple curves with increasing $|\nu|$. The curves are offset as indicated by the gray dashed lines. Stars indicate the center of the regions identified with electron bubble states.  
(c) Temperature dependence of electron bubble states in $N=3$ and (d) $N=4$ LLs, measured at $B=13T$.  
}
\end{figure}

For $N\geq2$ (Figs. \ref{fig:figure1}c-e), a qualitatively different behavior is observed, with $\mu$ dominated by much weaker oscillatory features that are not associated with any particular fractional $\nu$. 
As we elaborate upon below, these features are signatures of multi-electron bubble states. Bubble states are generically expected in higher LLs due to the nature of the single-particle wave functions, which feature multiple nodes.  This form factor considerably modifies the Coulomb repulsion at short distances, favoring charge-density-wave-type states instead of incompressible fractional quantum Hall states. In the $N=2$ LL, our measurement reveals a competition between the FQH states observed at $\nu^*=-1/5$ and $-4/5$ and electron bubble states, as reported previously\cite{chen_competing_2019}. 
In the $N=3$ and $N=4$ LLs, the electron bubble phases are favored over the entire range of filling factors, manifesting as a slow modulation of $\mu$ and $d\mu/d\nu$, as shown in \ref{fig:figure2}a-b. The number of oscillatory features increases with $N$.  In the $N=3$ and $N=4$ LL we observed three and four pairs of features, related by particle hole symmetry about $\nu^*=1/2$, respectively.

The panels of Fig. \ref{fig:figure2}a-b show $d\mu/d\nu$ measured over a range spanning several LLs each, grouped by their orbital quantum number.  For the $N=3$ orbital (Fig. \ref{fig:figure2}a), the four curves depicted are acquired in filling factor ranges corresponding to each of the four symmetry broken levels spanning $-10<\nu<-6$.  Due to limitations on the range of the electrostatic gates, for the $N=4$ LL (Fig. \ref{fig:figure2}b) only $-12<\nu<-10$ is shown. Remarkably, the repetition of the pattern of $\mu$ oscillations across different symmetry-broken levels indicates that this physics is independent of the spin and valley order. 
We may conclude that the formation of the bubble phases is governed only by single-component LL physics; as a consequence, the bubbles are not expected to be accompanied by complex spin or valley textures as have been shown to play a role in lower LLs\cite{zhou_solids_2019,liu_visualizing_2022}.

The energy scale characterizing the bubble phases may be directly accessed via the temperature dependence, shown in Fig.  \ref{fig:figure2}c-d.  Signatures of the bubble phases disappear rapidly for $T\approx 1-2K$ in the $N=3$ LL, and below 1K in the $N=4$.  
This is consistent with the general scale of the chemical potential changes associated with these phases, which are on the order of a few hundred $\mu eV$, as well as previously reported transport data\cite{chen_competing_2019}. 
The order of magnitude of this scale is consistent with  simplified Lindemann criterion\cite{lindemann_calculation_1910} for crystal melting, according to which the thermal position fluctuations need to be roughly $15\%$ of the lattice spacing to make the crystal melt. Within the harmonic approximation for the crystal, one obtains critical temperatures in the $\sim 1$ K range (see supplementary material). Notably, the energy scale of the bubble phases is considerably smaller than that of the fractional quantum Hall physics in the lower LLs, where gaps (at comparable magnetic fields) typically are in the $>10K$ range.

\begin{figure}[t!]
\includegraphics{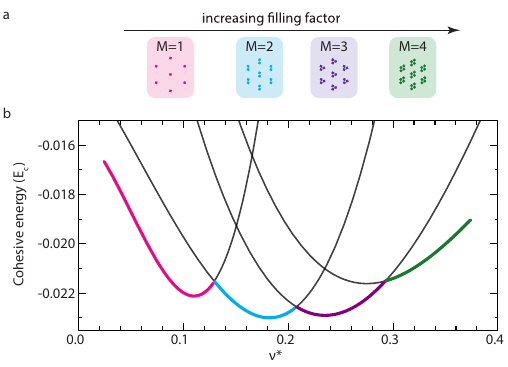}
\caption{\label{fig:figure3} \textbf{Cohesive energy for electron bubble states.}
(a) Schematic depiction of electron bubble phases in the $N=4$ LL. 
(b) Calculated cohesive energy for the $N=4$ LL (see supplementary information for details). The ground state is obtained by tracing the lowest energy state at each filling factor, which is highlighted by colored lines. The color codes here match those in panel (a).
}
\end{figure}

Theoretically, the ground state of the interacting electron system in a partially filled high-N Landau level is expected to evolve through a series of multi-electron bubble phases, as illustrated in Fig. \ref{fig:figure3}a for the case of N=4. 
These crystalline phases can be  described within a mean-field approach as presented in detail in the Supplementary Material.
Fig. \ref{fig:figure3}a shows the cohesive energy per particle for the bubble crystals with $M$ electrons per lattice site as a function of the effective filling factor $\nu^*$. The cohesive energy is the energy per particle, from which we have already subtracted the Hartree-Fock energy of a featureless electronic liquid\cite{goerbig_competition_2004} as well as the charging energy of the parallel plate capacitor in which the sample is embedded. 
For a fixed value of $M$, the energy of the triangular bubble crystals depends on the spacing $\Lambda_B=\sqrt{4\pi M/\sqrt{3}\nu^*}l_B$ between the bubbles, which in turn depends on the effective filling $\nu^*$. Here, $l_B=\sqrt{\hbar/eB}$ is the magnetic length. 

One obtains a family of curves, with minima at positions described approximately by  $\nu^*\sim M/N$.  The $M$-bubble phase is realized whenever it is lowest in energy within a a certain filling-factor range. Within a given Landau level, the maximum stabilized value of $M$ equals $N$. 
Theoretically, one may even stabilize a bubble phase with $M=N+1$ in the vicinity of a half-filled, singly-degenerate Landau level ($\nu^*\sim 1/2$)\cite{cote_dynamics_2003}. However, this phase is thought to compete energetically with a stripe phase; we find no evidence for it in the experimental data.  

Notably, the family of minimum energy curves shown in Fig. \ref{fig:figure3}b are not convex upon variation of $\nu^*$, a signature of thermodynamic instability to the formation of mixed phases in which parts of the sample area are occupied by crystals with differing number of electrons per bubble.  
However, we note that for our experimental geometry, the variations in internal energy caused by the bubble phases are dwarfed by the electrostatic energy of the electron gas.
Taking this into account, mixed phases are only found in a range $\delta \nu\approx 2\times10^{-3}$ (see supplementary information) in the vicinity of the level crossings visible in Fig. \ref{fig:figure3}b.  
In this picture, then, we expect a succession of pure bubble phases, separated by sharp phase transitions.

\begin{figure*}[ht!]
\includegraphics{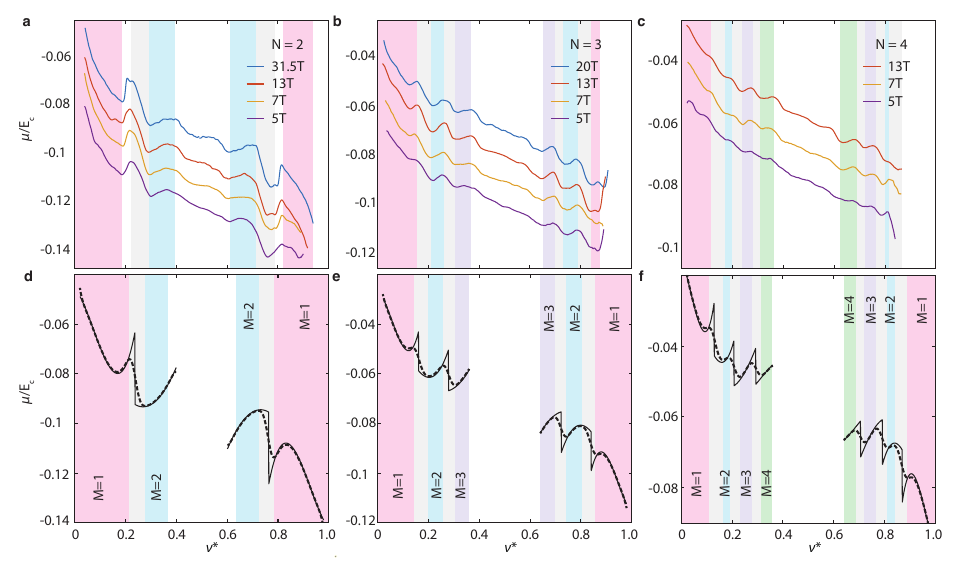}
\caption{\label{fig:figure4} \textbf{Quantitative comparison with theoretical model of electron bubble cascade}. 
(a) $\mu(\nu)$ at several magnetic fields in the $N=2$, (b) $N=3$, and (c) $N=4$ Landau level. 
The data at $B=31.5T$ and $20T$ were measured at $300mK$, while the data at $13T$, $7T$, and $5T$ were measured at $15mK$. The chemical potential change is presented in units of the Coulomb energy 
$E_c=\frac{e^2}{\epsilon l_B} \approx 12.5 meV\cdot\sqrt{B/\mathrm{Tesla}}$. 
The red, orange, and purple curves are offset by $-0.01E_c$, $-0.02E_c$, and $-0.03E_c$ from the blue curve, respectively. 
(d) Chemical potential calculated by mean field-theory (solid lines, see supplementary materials) in the $N=2$, (e) $N=3$, and (f) $N=4$ Landau level. The dashed lines in these panels are chemical potential taking disorder broadening into account. The pink, blue, purple, and green color bars represent the domain of stability for the $M=1$, $M=2$, $M=3$, and $M=4$ electron bubble phases within the disorder broadened model, respectively. The gray regions represent broadened phase transitions where neighboring pure electron bubble phases coexist. Panel (a)-(c) use the same color codes to label the corresponding regions identified by experiments from the sign of the compressibility. 
}
\end{figure*}

To facilitate comparison between experiment and theory, in Fig.~\ref{fig:figure4}a-c, we plot the experimentally measured $\mu$ scaled by the Coulomb energy, $E_c=e^2/(\epsilon\ell_B)$. Each panel presents $\mu$  measured at different values of the magnetic field $B$ for the same LL fillings, with an offset of $0.01E_C$ between curves introduced for clarity. The $\mu$ modulations observed in the curves are almost identical in these units, as expected given the Coulomb-driven nature of the electron bubble phases.
Fig.~\ref{fig:figure4}d-f presents the calculated chemical potential of electron bubble phases in the $N=2$, $N=3$, and $N=4$ LLs in the absence of disorder. 
The solid curves are obtained from the calculated energy per particle $E$ of the $M$-bubble phases via $\mu=\partial \left(\nu E\right)/\partial \nu$\cite{goerbig_competition_2004}. 
Note that in these calculations, we restore the contribution of the featureless background charge omitted above in the calculation of the cohesive energy. Our calculations account for screening caused by both the dielectric environment as well as inter-Landau level excitations in the graphene\cite{shizuya_electromagnetic_2007,roldan_magnetic_2010}. 
As in the N=0 and N=1 Landau levels\cite{yang_experimental_2021}, accurately accounting for screening is required for quantitative agreement between experiment and theory in graphene.

Despite the comparative simplicity of our model, it agrees quantitatively with the data in the overall scale of the chemical potential modulation across the Landau level, as well as in the locations of the various bubble phases, which we identify with positive compressibility regions for the $M>=2$. 
However, in contrast to the theoretical model, where the phase transitions are sharp, in the experimental data the phase transitions are marked by broad regions of negative compressibility typically rather than sharp jumps.  
It is natural to associate these regions with a mixed phase arising from disorder potentials.  To capture this physics, we convolve the disorder-free curves with a Gaussian `inhomogenous broadening' of width $\Delta\nu=0.015$ at $13T$. Given the negligible quantum capacitance in the bubble regime, this is equivalent to an energy broadening $\Delta E=7.5meV$.  
The dashed curves in Fig. \ref{fig:figure4}d-f show the results of this model. We use the same color code to label the regions associated with pure and mixed electron bubble phases in both experimental and simulation data in the figure; the disordered model quantitatively reproduces the key missing feature of the experimental data, replacing the cusps of the disorder-free model with negative compressibility regimes as observed experimentally.

We note in closing several open questions raised by our work.  First, while electron solids evidently dominate the ground states for $N>2$, it is likely that they appear in the lower LLs as well, but are difficult to detect with bulk methods where their subtle thermodynamic or transport phenomenology may be overwhelmed by the incompressibility of the fractional quantum Hall states.  Second, it is unclear whether the particular orbital wavefunctions of single- and multi-layer graphene may lead to any particularities in the electron solid ground states as compared to semiconductor systems.  Finally, our disorder model is likely to be gross oversimplification.  In particular, the lack of observed magnetic field dependence in the sharpness of the phase transitions is at odds with a model of quenched disorder where the effective broadening $\Delta E$ would be expected to be magnet field independent.  
These and other questions might be directly resolved via scanning tunneling microscopy measurements of the real space structure of these phases\cite{coissard_imaging_2022,liu_visualizing_2022}, as well as more detailed theoretical modeling that accounts for the interplay of disorder, finite temperature, and mesoscopic phase separation. 

\begin{acknowledgments}
The authors acknowledge discussions with M. Zaletel. 
This work was primarily supported by Office of Naval Research under award N00014-23-1-2066. 
A.F.Y. acknowledges the additional support of the Gordon and Betty Moore Foundation EPIQS program under award GBMF9471. 
A portion of this work was performed at the National High Magnetic Field Laboratory, which is supported by National Science Foundation Cooperative Agreement No. DMR-1644779 and the State of Florida.  
This work made use of shared facilities supported by 
the National Science Foundation through Enabling Quantum Leap: Convergent Accelerated Discovery Foundries for Quantum Materials Science, Engineering and Information (Q-AMASE-i) award number DMR-1906325. 
K.W. and T.T. acknowledge support from JSPS KAKENHI (Grant Numbers 19H05790, 20H00354 and 21H05233).
S.J.\ and B.S.\ were supported by the NSF under Grant
No.\ DMR-2045742.
   
\end{acknowledgments}

\clearpage
\pagebreak
\widetext
\begin{center}
\textbf{\large Supplementary Information}
\end{center}
\renewcommand{\thefigure}{S\arabic{figure}}
\renewcommand{\thesubsection}{S\arabic{subsection}}
\setcounter{secnumdepth}{2}
\renewcommand{\theequation}{S\arabic{equation}}
\renewcommand{\thetable}{S\arabic{table}}

\setcounter{figure}{0}
\setcounter{equation}{0}
\setcounter{page}{1}
\onecolumngrid

\section{Model of electron bubble states in N$>=$2 LL}
\subsection{Energy of the bubble phases}

The energy of the bubble crystals with $M$ electrons per bubble at a partial filling $\nu^*$ is readily calculated with the help of the formula\cite{goerbig_competition_2004}
\begin{equation}\label{S1}
    E^M_{coh}(\nu^*)=\frac{\nu^*}{2{\pi}l^2_BM}\sum_{\mu_1,\mu_2}u^{HF}_N(G_{\mu_1,\mu_2})\frac{J^2_1(\sqrt{2M}|G_{\mu_1,\mu_2}|)}{|G_{\mu_1,\mu_2}|^2}
\end{equation}
where $l_B$ is the magnetic length, and $G_{\mu_1,\mu_2}={\mu_1}\mathbf{g_1}+{\mu_2}\mathbf{g_2}$ are reciprocal lattice vectors, with $\mathbf{g_1}=2{\pi}(\mathbf{e}_x-\mathbf{e}_y/\sqrt{3})/{\Lambda}_B$ and $\mathbf{g_2}=4{\pi}\mathbf{e}_y/\sqrt{3})/{\Lambda}_B$, where
\begin{equation}\label{eq:bubble_space}
    {\Lambda}_B=\sqrt{\frac{4{\pi}M}{\sqrt{3}\nu^*}}l_B
\end{equation}
is the lattice spacing. Furthermore,
\begin{equation}\label{S3}
    u^{HF}_N(\mathbf{q})=v_N(\mathbf{q})-\frac{2{\pi}l^2_B}{A}\sum_\mathbf{p}v_N(\mathbf{p})e^{-i(p_xq_y-p_yq_x)l^2_B}
\end{equation}
is the Hartree-Fock potential in the N-th LL, with
\begin{equation}\label{S4}
    v_N(\mathbf{q})=\frac{2{\pi}e^2}{{\epsilon}q}|F_N(ql_B)|^2
\end{equation}
being the effective interaction in this level, in terms of the LL form factor ($N\geq1$)
\begin{equation}
    F_N(\mathbf{q})=\frac{1}{2}\left[L_N\left(\frac{x}{2}\right)+L_{N-1}\left(\frac{x}{2}\right)\right]e^{-x/4}
\end{equation}
which takes into account the overlap between the LL wave functions, in terms of Laguerre polynomials $L_N(x)$. Furthermore, we have used the measure $\sum_{\mathbf{q}}=[A/(2\pi)^2]{\int}d^2q$, where $A$ is the total area. Notice that the cohesive energy is an energy per particle, where the reference energy is that of the uniform electron gas, where we do not only take into account its exchange energy
\begin{equation}\label{S6}
\begin{split}
    E_{ref}(\nu^*)&=-\frac{\nu^*}{4{\pi}l^2_B}u^{HF}_N(\mathbf{q}=0)=-\frac{\nu^*}{2A}\sum_{\mathbf{p}}v_N(\mathbf{p})\\&=-\frac{\nu^*}{2}\left(\frac{e^2}{{\epsilon}l_B}\right){\int}^{\infty}_0dx[F_N(x)]^2
\end{split}
\end{equation}
where $x=ql_B$. In order to calculate the chemical potential of the system, one needs to take into account also this contribution,
\begin{equation}\label{S7}
\begin{split}
    {\mu}_{\nu^*}&=\frac{\partial[\nu^*(E^M_{coh}+E_{ref})]}{{\partial}{\nu^*}}\\&=E^M_{coh}+\nu^*\frac{{\partial}E^M_{coh}}{{\partial}{\nu^*}}-\nu^*\left(\frac{e^2}{{\epsilon}l_B}\right){\int}^{\infty}_0dx[F_N(x)]^2
\end{split}
\end{equation}
The integral in the last term is given by
\begin{equation}
   {\int}^{\infty}_0dx[F_N(x)]^2=\frac{515}{1024}\sqrt{\frac{\pi}{2}}=0.630
\end{equation}
Notice that the last term in Eq.~\eqref{S7} can be viewed, similarly to the reference energy \eqref{S6}, as a reference chemical potential $\mu_{ref}(\nu^*)$ that is due to the underlying structure-less electron liquid, and one has 
\begin{equation}
   \mu_{ref}(\nu^*)=2E_{ref}(\nu^*)
\end{equation}

\begin{figure*}[ht!]
\includegraphics{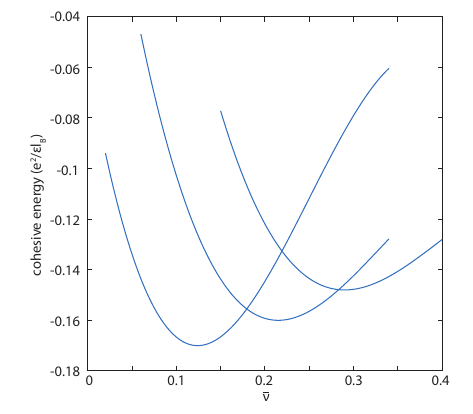}
\caption{\label{fig:SI_cohesiveEnergy}Theoretical calculation of the cohesive energy for electron bubble phases in the $N=3$ Landau level. The blue lines represent the cohesive energy of M-electron bubble phases.}
\end{figure*}

\subsection{Energy and chemical potential of the bubble phases for a screened interaction potential in the $N=3$ Landau level}
The above calculations do not taken into account the role of (non-local) screening, but only the dielectric environment in the form of the overall dielectric constant $\epsilon=4.5$ (for BN-encapsulated graphene). In principle the screening of the Coulomb energy due to the metallic top and bottom gates at distance $d$ from the graphene also needs to be taken into account. If we have a top gate at distance $d_1$ and a bottom gate at distance $d_2$, the $2{\pi}e^2/{\epsilon}q$ Coulomb potential needs to be multiplied by a "form factor"
\begin{equation}
    f_{d_1,d_2}(q)=2\frac{\tanh(d_1q)\tanh(d_2q)}{\tanh(d_1q)+\tanh(d_2q)}
\end{equation}
which reduces to $f_{d,d}(q)=\tanh(dq)$ in the case of a symmetric arrangement of the gates, with $d=d_1=d_2$.
However, in the present experiment, we can safely neglect the effect of the gate screening, which are at distances much larger compared with the magnetic length (and the cyclotron radius), so that $dq\sim d/l_B\gg1$, in which limits $f_{d_1,d_2}\simeq1$.

\begin{figure*}[ht!]
\includegraphics{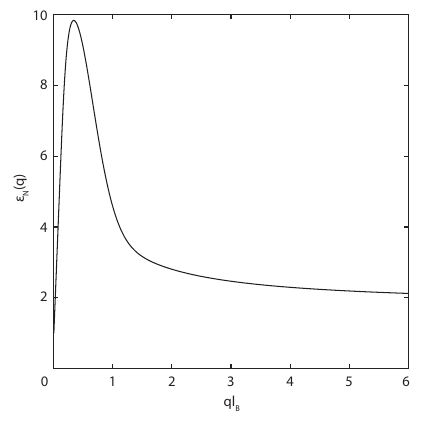}
\caption{\label{fig:SI_RPA}Static dielectric function for $N = 3$ as a function of the wave vector, calculated from Eq.\eqref{S16}. 
}
\end{figure*}

In contrast to the gate screening, one needs to take into account the screening due to virtual excitations to Landau levels adjacent to $N=3$. This is captured in the (static) dielectric function, which one obtaines from an RPA calculation
\begin{equation}
    v^{scr}_N(\mathbf{q})=\frac{2{\pi}e^2}{\epsilon\epsilon_N(q)q}|F_N(nl)B)|^2
\end{equation}
in terms of the polarizability $\Pi^0_n(q,\omega)$ for non-interacting electrons in a magnetic field\cite{shizuya_electromagnetic_2007,roldan_magnetic_2010}. The polarizability consists of a sum over all inter-Landau-level transitions allowed by the Pauli principle and has a particular form due to the presence of inter-band excitations from the valence to the conduction band. In contrast to the polarizability of the usual 2D electron gas, it is not cut off above twice the Fermi wave vactor $k_F\simeq\sqrt{2n}/l_B$, but continues to grow linearly. This linearity provides us with a contribution to the dielectric constant in the large wave-vector limit
\begin{equation}
    \epsilon_n(q\rightarrow\infty)=1+\pi\alpha_G/2+\alpha_Gk_F/q
\end{equation}
where $\alpha_G=e^2/{\hbar}v_F\epsilon=2.2/\epsilon\simeq0.49$ is the graphene fine-structure constant\cite{roldan_magnetic_2010}. Furthermore, the asymptotic behavior in the large-$q$ limit shows that one retrieves a simple Thomas-Fermi screening on top of this constant background, as indicated by the last term in the above equation. In th opposite small wave-vector limit, the static dielectric function behaves as 
\begin{equation}
    \epsilon_n(q\rightarrow0)=1+2\pi\alpha_GN^{3/2}ql_B
\end{equation}
while one has a maximum that scales as $\epsilon_n(q\sim1/R_C)\sim\alpha_Gn$, where $R_C=\sqrt{2N}l_B$ is the graphene cyclotron radius. A convenient interpolation formula that reproduces well the static dielectric function is given by
\begin{equation}\label{S16}
\begin{split}
    \epsilon_n(q)=1+2\pi\alpha_GN\left[\frac{qR_C}{\sqrt{2}}e^{-q^2R_C^2/2}+\left(q^2l_B^2+\frac{2\sqrt{2}N}{\pi}ql_B\right)\left(1-e^{-1/(2q^2R_C^2)}\right)\right]
\end{split}
\end{equation}
The form of the static dielectric function for the Landau level $N=3$ is shown in Fig.~\ref{fig:SI_RPA}.

The screened interaction potential is now obtained by dividing Eq.~\ref{S4} by $\epsilon_n(q)$
\begin{equation}
    v^{scr}_N(\mathbf{q})=\frac{2{\pi}e^2}{{\epsilon}{\epsilon}_N(q)q}|F_N(ql_B)|^2
\end{equation}
Naturally, this affects also the exchange as well as the Hatree-Fock potential Eq.~\ref{S3}. Finally, one notices that the reference energy \ref{S6} now reads
\begin{equation}
    E^{scr}_{ref}(\nu^*)=-\frac{\nu^*}{2}\left(\frac{e^2}{{\epsilon}l_B}\right){\int}^{\infty}_0dx\frac{[F_N(x)]^2}{\epsilon_N(x)}
\end{equation}
The integral yields the numerical value
\begin{equation}
{\int}^{\infty}_0dx\frac{[F_N(x)]^2}{\epsilon_N(x)}=0.143
\end{equation}
and one notices here already a reduction by a factor of 4.5. One thus obtains the reference energy
\begin{equation}
E^{scr}_{ref}(\nu^*)=-0.0715\nu^*\left(\frac{e^2}{{\epsilon}l_B}\right)
\end{equation}
and the reference chemical potential of the uncorrelated liquid state.
\begin{equation}
\mu^{scr}_{ref}(\nu^*)=-0.143\nu^*\left(\frac{e^2}{{\epsilon}l_B}\right)
\end{equation}

\begin{figure*}[ht!]
\includegraphics{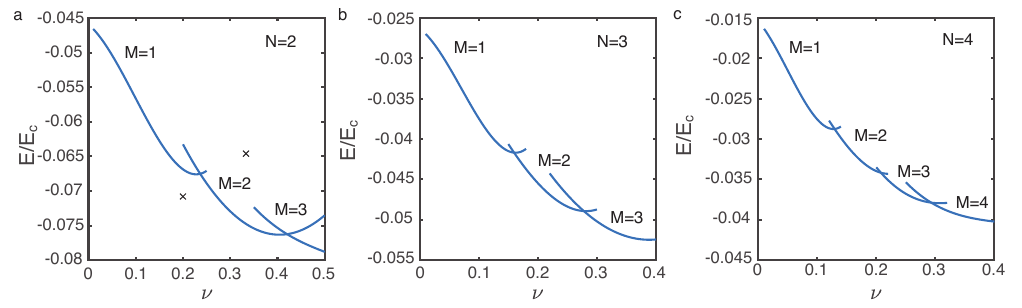}
\caption{\label{fig:SI_bubbles}Theoretical calculation of energy per electrons for electron bubble phases in the $N=2$(a), $N=3$(b), and $N=4$(c) LLs. The blue lines represent the energy of M-electron bubble phases. In the $N=2$ LL, we also plot the energy of the Laughlin states for partial filling factor $\nu^*=1/3$ and $\nu^*=1/5$. While at $\nu^*=1/3$ the electron bubble phase has lower energy, at $\nu^*=1/5$ the FQH state is more stable, which indeed agree with the experimental observation. Around half filling, previous calculation in GaAs shows the unidirectional CDW phase dominates\cite{goerbig_possible_2004}, which we didn't consider here.}
\end{figure*}

The cohesive energies are calculated by numerical integration since there is no compact analytical formula for the exchange potential in the case of the screened interaction potential. The energy of the cohesive energy for the $N=3$ LL are given by the fitted formulas (to order three in the filling factor measured from the energy minimum; Fig.~\ref{fig:SI_bubbles}b.)
\begin{equation}
\frac{E^{scr,M=1}_{coh}(\nu^*)}{e^2/{\epsilon}l_B}=-0.0311+0.952(\nu^*-0.132)^2+5.14(\nu^*-0.132)^3
\end{equation}

\begin{equation}
\frac{E^{scr,M=2}_{coh}(\nu^*)}{e^2/{\epsilon}l_B}=-0.0312+0.412(\nu^*-0.21)^2+0.812(\nu^*-0.21)^3
\end{equation}
and
\begin{equation}
\frac{E^{scr,M=3}_{coh}(\nu^*)}{e^2/{\epsilon}l_B}=-0.0291+0.287(\nu^*-0.265)^2+0.014(\nu^*-0.265)^3
\end{equation}

\subsection{Energy and chemical potential of the $N=2$ bubble phases for a screened interaction potential}

In the $N=2$ LL, the FQH states are in competition with the electron bubble states. Here we show the calculation of the energy of the Laughlin states at $\nu^*=1/3$ and $\nu^*=1/5$. Their cohesive energy can be calculated by the Haldane's pseudo-potential
\begin{equation}
    \frac{V^N_{2m+1}}{e^2/{\epsilon}l_B}=2{\pi}{\int}^{\infty}_0dxxv^{scr}_N(x)L_{2m+1}(x^2)e^{-x^2/2}
\end{equation}
and the cohesive energy
\begin{equation}
    E^L_{coh}(\nu^*=1/(2s+1))=\frac{\nu}{\pi}\sum^{\infty}_{m=0}c^s_{2m+1}V^N_{2m+1}
\end{equation}
The coefficients $c^s_{2m+1}$ characterize the Laughlin state at $\nu^*=1/(2s+1)$ and can be calculated, within a good approximation, with the help of sum rules from the plasma analogy\cite{goerbig_competition_2004}. With this method and taking into account the reference energy
\begin{equation}
    E^{scr}_{ref}(\nu^*)=-0.1\nu^*\left(\frac{e^2}{{\epsilon}l_B}\right)
\end{equation}
One obtains
\begin{equation}
    E^L(\nu^*=1/3)=-0.0646\left(\frac{e^2}{{\epsilon}l_B}\right)
\end{equation}
and
\begin{equation}
    E^L(\nu^*=1/5)=-0.0708\left(\frac{e^2}{{\epsilon}l_B}\right)
\end{equation}
for the energy per particle of the Laughlin states. We notice that only the quantum liquid at $\nu^*=1/5$ is stable, while the $1/3$-state has a higher energy than the $M=2$ bubble phase at the same filling (Fig.~\ref{fig:SI_bubbles}a, black crosses).

The cohesive energies of the pure bubble crystals now read
\begin{equation}
\frac{E^{scr,M=1}_{coh}(\nu^*)}{e^2/{\epsilon}l_B}=-0.0478+0.478(\nu^*-0.16)^2+2.53(\nu^*-0.16)^3
\end{equation}

\begin{equation}
\frac{E^{scr,M=2}_{coh}(\nu^*)}{e^2/{\epsilon}l_B}=-0.0439+0.319(\nu^*-0.245)^2-0.0217(\nu^*-0.245)^3
\end{equation}

and
\begin{equation}
\frac{E^{scr,M=3}_{coh}(\nu^*)}{e^2/{\epsilon}l_B}=-0.038+0.347(\nu^*-0.305)^2-0.539(\nu^*-0.305)^3
\end{equation}

The $M=3$ bubble crystal in the vicinity of $\nu^*=1/2$ is in competition with a stripe phase or a highly anisotropic Wigner crystal, and therefore might not exist as the ground state.

\subsection{Energy and chemical potential of the $N=4$ bubble phases for a screened interaction potential}

Finally we show the calculated energy per particle of the bubble phases and the associated chemical potentials in the $N=4$ LL (Fig.~\ref{fig:SI_bubbles}c). As expected from the scaling $M_{max}=N$ of the bubble crystal with the maximum number of particles per bubble in the $N$-th LL, we need to consider a bubble crystal with four electrons per bubble now. (Notice that the originally proposed $M_{max}=N+1$ bubble crystal is usually covered by the stripe or the anisotropic Wigner crystal phase around half filling). In the $N=4$ LL, the reference energy is given by
\begin{equation}
    E^{scr}_{ref}(\nu^*)=-0.0559\nu^*\left(\frac{e^2}{{\epsilon}l_B}\right)= \frac{\mu^{scr}_{ref}(\nu^*)}{2}
\end{equation}
while the cohesive energies of the pure bubble crystals read
\begin{equation}
\frac{E^{scr,M=1}_{coh}(\nu^*)}{e^2/{\epsilon}l_B}=-0.0221+1.409(\nu^*-0.11)^2+7.77(\nu^*-0.11)^3
\end{equation}

\begin{equation}
\frac{E^{scr,M=2}_{coh}(\nu^*)}{e^2/{\epsilon}l_B}=-0.023+0.625(\nu^*-0.182)^2+1.69(\nu^*-0.182)^3
\end{equation}

\begin{equation}
\frac{E^{scr,M=3}_{coh}(\nu^*)}{e^2/{\epsilon}l_B}=-0.0229+0.454(\nu^*-0.235)^2-0.635(\nu^*-0.235)^3
\end{equation}

and
\begin{equation}
\frac{E^{scr,M=4}_{coh}(\nu^*)}{e^2/{\epsilon}l_B}=-0.0216+0.342(\nu^*-0.275)^2-0.823(\nu^*-0.275)^3
\end{equation}

\section{Theoretical estimate for the range of phase coexistence}

Here we consider the general criteria for phase coexistence between two distinct electronic phases at zero temperature and in the absence of disorder, and we estimate the maximum range of filling factor $\delta \nu$ that such a coexistence regime can occupy. 

Consider a phase transition between two distinct phases of the electron system, which we label as 1 and 2. If $E_{i}(n)$ denotes the energy per electron as a function of the electron density $n$ in phase $i = 1,2$, then at zero temperature the free energy per unit area in phase $i$ is $f_i(n) = n E_i(n)$.  In the usual Maxwell construction for phase coexistence, one considers that the total free energy for a fixed number of electrons is minimized with respect to the area fraction $x$ that is occupied by one of the two phases (say, phase 2). That is, if $n_1 , n_2$ denote the electron densities in spatial regions occupied by phases $1$ and $2$ respectively, then the region of phase coexistence can be found by minimizing the total free energy $(1-x)f_1(n_1) + x f_2(n_2)$ with respect to $x$, $n_1$, and $n_2$ for a fixed average density $\bar{n} = (1-x)n_1 + x n_2$. From this procedure one can easily show that, within the regime of phase coexistence, the two phases have the same chemical potential $\mu_i = df_i(n) /dn$.

In 2D electron systems, however, there is an additional Coulombic energy cost associated with the difference between the local electron density ($n_1$ or $n_2$) and the charge density of the positive background (which for large gate separation can be treated as spatially uniform, with charge density $+e\bar{n}$), which leaves one phase with a net positive charge and the other with a net negative charge. This Coulombic energy suppresses phase coexistence, and generally prohibits the formation of macroscopic phase domains (since a region of size $L$ and charge density $\eta$ has a Coulombic self-energy per unit area that grows extensively with the region size, $\sim \eta^2 L$). Thus, phase coexistence, where present, can only take the form of ``microemulsion''-type phases \cite{spivak_phases_2004, spivak_colloquium_2010}, in which stripes or droplets of the less abundant phase are interspersed periodically among the more abundant phase.

To derive the criteria for phase coexistence including this Coulomb energy cost, one should minimize the total free energy per unit area
\be 
f=\left(1-x\right)f_{1}\left(n_{1}\right)+xf_{2}\left(n_{2}\right)+e_{m},
\label{eq:f}
\ee
where $e_{m}$ is the energy density of mixing, which contains both electrostatic and surface energy terms. 
The energy $e_{m}$ has been considered in detail in Ref.~\cite{ortix_frustrated_2006} for both the droplet and the stripe geometry. For the case of droplet configurations, $e_m$ is well-approximated by
\be 
e_m = \frac{8}{\sqrt{3}} |n_2 - n_1| \sqrt{\frac{e^2 \gamma}{\epsilon}} x (1-x),
\ee 
where $\gamma$ is the surface tension. (In stripe configurations there is an additional logarithmic factor \cite{ortix_frustrated_2006}.) The corresponding optimal droplet size $R_d$ is
\be 
R_d = \frac{\sqrt{3}}{2\sqrt{x\left(1-x\right)}} \frac{\sqrt{e^2 \gamma/ \epsilon }}{\left(e^2/\epsilon\right)|n_2 - n_1|} . 
\label{eq:dropletR}
\ee 
Thus, both the droplet size and the energy of mixing depend on the parameter $\beta \equiv \sqrt{e^2 \gamma / \epsilon}$, which has units of energy. While in the usual Maxwell construction the two phases have the same chemical potential, the inclusion of the term $e_m$ implies that two coexisting phases have a difference in chemical potential $\approx 8 \beta/ \sqrt{3}$.

This approach gives two primary criteria for the existence of a mixed phase, which ultimately place a constraint on the range of average density $\bar{n}$ that such a mixed phase may occupy. First, the difference in chemical potential between the two phases must be equal to $8 \beta/\sqrt{3}$.  The chemical potential difference is maximal when the two phases have a density nearly equal to the value $n_c$ at which the two phases have the same energy, $E_1(n_c) = E_2(n_c)$.  Correspondingly phase coexistence is only permitted when
\be 
\beta < \frac{\sqrt{3}}{8} \left| \mu_2(n_c) - \mu_1(n_c) \right|.
\label{eq:coexist1}
\ee 
So phase coexistence requires the parameter $\beta$ to be small enough.

On the other hand, if $\beta$ is too small, then the length scale associated with microemulsion droplets becomes unrealistically small [see Eq.~(\ref{eq:dropletR})]. Specifically, phase coexistence in the sense of mesoscopic domains of two distinct phases is not possible unless the corresponding droplet size is much larger than the inter-electron spacing: $R_d \gg n_{1,2}^{-1/2}$. This condition sets a \emph{lower} bound on the parameter $\beta$:
\be 
\beta \gg \frac{e^2 |n_2 - n_1|}{\sqrt{3} \epsilon n_{1,2}^{1/2}}.
\label{eq:coexist2}
\ee

Combining Eqs.~(\ref{eq:coexist1}) and (\ref{eq:coexist2}) produces an inequality for the maximum difference in density between the two phases:
\be 
|n_2 - n_1| \ll \frac{3}{8} \frac{\left| \mu_2(n_c) - \mu_1(n_c) \right| n_{c}^{1/2}}{e^2/\epsilon}.
\ee 
Inserting units of the magnetic length, we arrive at
\be 
\delta \nu \ll \frac{3 \sqrt{2 \pi}}{8} \frac{\left| \delta \mu(\nu_c) \right| }{e^2/\epsilon l_B} \nu_c^{1/2},
\label{eq:deltanu}
\ee 
where $\delta \nu$ is the window of filling factor that corresponds to phase coexistence, $\nu_c$ is the filling factor at which the two phases have the same energy, and $\delta \mu(\nu_c)$ is the difference in chemical potential between the phases at $\nu = \nu_c$.

For the case depicted in Fig.~3 of the main text, which illustrates three phase transitions between distinct bubble phases, Eq.~(\ref{eq:deltanu}) gives $\delta \nu < 1.8 \times 10^{-3}$, $\delta \nu < 2.5 \times 10^{-3}$, and $\delta \nu < 2.8 \times 10^{-3}$ for the respective windows of phase coexistence.

\section{Lindemann criterion for the melting of the bubble phase}

In order to understand the orders of magnitude for the melting of the bubble crystal, one may appeal to the phenomenological Lindemann criterion \cite{lindemann_calculation_1910} according to which a crystal melts if the temperature-induced position fluctuations $\Delta x$ are fraction of the lattice spacing $\Lambda_B$. The thumb rule is given by \cite{lindemann_calculation_1910} 
\begin{equation}
    \Delta x = C \Lambda_B, \qquad  \text{with} \qquad C\simeq 0.15.
\end{equation}
Within the harmonic approximation, the potential-energy scale is given by 
\begin{equation}
    E_\text{pot}\simeq \frac{e^2}{\epsilon \Lambda_B}\left(\frac{\Delta x}{\Lambda_B}\right)^2,
\end{equation}
which needs to be identified with the thermal energy $k_BT$ so that one obtains for the critical temperature 
\begin{equation}
    k_BT^*\simeq C^2\frac{e^2}{\epsilon \Lambda_B}\simeq 
    C^2\sqrt{\frac{\sqrt{3}\nu^*}{4\pi M}}\left(\frac{e^2}{\epsilon l_B}\right),
\end{equation}
where we have made use of the expression (\ref{eq:bubble_space}) for the spacing in the bubble crystal. If we consider a Coulomb energy scale of 
\begin{equation}
    \frac{e^2}{k_B \epsilon l_B}\simeq 110 \times \sqrt{B[\text{T}]}\, \text{K},
\end{equation}
which is appropriate for graphene encapsulated in h-BN, this yields roughly 400 K at a magnetic field of $B=13$ T, as in our experiments (see Fig. 2 of the main text). For a filling factor $\nu^*=0.2$, where we expect an $M=2$ bubble phase in $N=4$, this yields a critical temperature $T^*\simeq 1$ K, in good agreement with our experimental findings. 

Notice, however, that the above arguments remain valid only on the level of orders of magnitude. Indeed, since the crystals do not consist of pointlike lattice sites but rather of complex bubbles with an internal structure, the melting mechanism in the bubble phase may be more complicated than it is suggested by the simple Lindemann criterion. Furthermore, the interaction is not given by a simple Coulomb potential, but it has strong corrections at short distances that are expected to affect the electron dynamics inside each bubble. 

\end{document}